\documentclass[3p,times]{elsarticle}

\usepackage{ecrc}

\usepackage[pdfpagemode=UseOutlines,bookmarksnumbered,bookmarksopen]{hyperref}

\volume{00}

\firstpage{1}

\journalname{Nuclear Physics A}

\runauth{ Tan Luo, Yayun He, Xin-Nian Wang and Yan Zhu}

\jid{npa}

\jnltitlelogo{Nuclear Physics A}

\usepackage{amssymb}
\usepackage{amsmath}   %%math formula
\biboptions{square,comma,numbers,sort&compress}

\usepackage[figuresright]{rotating}

\begin{document}

\begin{frontmatter}

%% Title, authors and addresses

%% use the tnoteref command within \title for footnotes;
%% use the tnotetext command for the associated footnote;
%% use the fnref command within \author or \address for footnotes;
%% use the fntext command for the associated footnote;
%% use the corref command within \author for corresponding author footnotes;
%% use the cortext command for the associated footnote;
%% use the ead command for the email address,
%% and the form \ead[url] for the home page:
%%
%% \title{Title\tnoteref{label1}}
%% \tnotetext[label1]{}
%% \author{Name\corref{cor1}\fnref{label2}}
%% \ead{email address}
%% \ead[url]{home page}
%% \fntext[label2]{}
%% \cortext[cor1]{}
%% \address{Address\fnref{label3}}
%% \fntext[label3]{}

\dochead{}
%% Use \dochead if there is an article header, e.g. \dochead{Short communication}
%% \dochead can also be used to include a conference title, if directed by the editors
%% e.g. \dochead{17th International Conference on Dynamical Processes in Excited States of Solids}

\title{Jet propagation within a Linearized Boltzmann Transport Model \footnote{\vspace{-0.5cm}Talk presented by T. Luo at HP2013}}

%% use optional labels to link authors explicitly to addresses:
%% \author[label1,label2]{<author name>}
%% \address[label1]{<address>}
%% \address[label2]{<address>}

\author[a1]{Tan Luo}
\author[a1]{Yayun He}
\author[a1,a2]{Xin-Nian Wang}
\author[a3]{Yan Zhu}

\address[a1]{Institute of Particle Physics and Key Laboratory of Quark and Lepton Physics (MOE),
Central China Normal University, Wuhan 430079, China}
\address[a2]{Nuclear Science Division Mailstop 70R0319, Lawrence Berkeley National Laboratory, Berkeley, California 94740, USA}
\address[a3]{Departamento de F\'{i}sica de Part\'{i}culas and IGFAE, Universidade de Santiago de Compostela,
E-15706 Santiago de Compostela, Galicia, Spain}

\begin{abstract}
%% Text of abstract
A Linear Boltzmann Transport (LBT) model has been developed for the study of jet propagation inside a quark-gluon plasma. Both leading and thermal recoiled partons are transported according to the Boltzmann equations to account for jet-induced medium excitations. In this talk, we present our study within the LBT model in which we implement the complete set of elastic parton scattering processes. We investigate elastic parton energy loss and their energy and length dependence. We further investigate elastic energy loss and transverse shape of reconstructed jets. Contributions from the recoiled thermal partons are found to have significant influences on the jet energy loss and transverse profile.
\end{abstract}

\begin{keyword}
Jet quenching, jet transport, parton energy loss, quark-gluon plasma
%% keywords here, in the form: keyword \sep keyword

%% PACS codes here, in the form: \PACS code \sep code

%% MSC codes here, in the form: \MSC code \sep code
%% or \MSC[2008] code \sep code (2000 is the default)

\end{keyword}

\end{frontmatter}

%%
%% Start line numbering here if you want
%%
% \linenumbers

%% main text
%\section{Introduction}
\section*{}
%\label{}
{\bf Introduction}. Interaction between energetic jet shower partons and thermal partons in a quark-gluon plasma (QGP) is expected to lead to jet quenching~\cite{Gyulassy:2003mc} in high-energy heavy-ion collisions. The observed jet quenching has been considered as one of the most striking phenomena observed in heavy-ion collisions at the Relativistic Heavy-ion Collider (RHIC)~\cite{Adcox:2001jp,Adler:2002xw} and at the Large Hadron Collider (LHC)~\cite{LHC,LHCII}. The jet shower partons produced at the very early stage of a heavy-ion collision will suffer both collisional energy loss in the elastic $2\rightarrow 2$ scattering processes and radiative energy loss through induced gluon radiation. Induced gluon radiation has been considered as the dominant source of the parton energy loss for an energetic parton. However, many other studies also pointed out that the energy carried away by the recoiled partons in the binary elastic scattering could not be neglected \cite{Thoma1,DuttMazumder,Djordjevic,Ruppert} in order to account for the observed pattern of jet quenching in high-energy heavy-ion collisions at RHIC and LHC. The estimate of the elastic energy loss of a quark in the hot dense medium was first made by Bjorken \cite{BJ82}, and detailed studies were later carried out within the framework of finite temperature QCD \cite{THOMA351}. In addition to the energy transfer through elastic scattering, one should also consider change of flavor for the propagating parton which can only be taken into account systematically in a Monte Carlo transport model. We have therefore implemented the complete set of $2 \rightarrow 2$ elastic processes in QCD in the Linearized Boltzmann Transport (LBT) Monte Carlo model.

At the partonic level, jets defined by a jet-finding algorithm are composed by collimated showers of partons inside a jet cone of radius $R=\sqrt{(\phi-\phi_C)^2+(\eta-\eta_{C})^2}$. Previous work \cite{Wang:2013cia} within the Linearized Boltzmann Transport model \cite{Li:2010ts} with small angle approximation for all parton elastic scattering processes has shown that inclusion of the recoiled medium partons has a significant influence on the energy loss of reconstructed jets. For a more accurate and complete description, however, one needs to implement the complete set of elastic scattering processes including flavor changing annihilation and creation processes. We present here test simulations of jet propagation and jet-induced medium excitation within a static and homogeneous quark gluon plasma, and their effects on reconstructed jets. We consider only elastic processes here while inclusion of inelastic processes is still in development. A modified version of the anti-$k_t$ algorithm in FASTJET \cite{Cacciari:2011ma} is used to reconstruct the leading jet. The influence on the jet energy loss and jet transverse profile by the recoiled thermal partons will be studied in detail.

%\section{The model}
%\label{}

{\bf The model}. Simulations of interaction among leading, recoiled and thermal partons in LBT model are based on the Boltzmann transport equation,
\begin{equation}
\label{LBT}
p_1\cdot\partial f_1(p_1) = -\frac{1}{2}\int \negthickspace \frac{d^3p_2}{(2\pi)^3 2E_2} \negthickspace \int \negthickspace
\frac{d^3p_3}{(2\pi)^3 2E_3} \int \negthickspace \frac{d^3p_4}{(2\pi)^3 2E_4} \\
(f_1f_2-f_3f_4) \, |M_{12\rightarrow34}|^2 \, S_2(s, t, u) \\
(2\pi)^4\delta^4(p_1+p_2-p_3-p_4),
\end{equation}
for partonic processes $1+2\rightarrow 3+4$, where the matrix elements $|M_{12\rightarrow34}|^2$ are given by pQCD in terms of standard Mandelstam variables. As a  jet shower parton 1 traverses the QGP, it can scatter with parton 2 (light quark, antiquark or gluon) sampled from the medium. The parton phase-space distributions in a thermal medium with local temperature $T$ and the fluid velocity $u=(1, \vec{v})/\sqrt{1-\vec{v}^2}$ are denoted as $f_{i=2,4}(p_i)$, which are Bose-Einstein distributions for gluons and Fermi-Dirac distributions for quarks and antiquarks. The phase-space densities for the jet shower partons before and after scattering assume the form of a point-like particle$f_i=(2\pi)^3\delta^3(\vec{p}-\vec{p_i})\delta^3(\vec{x}-\vec{x_i}-\vec{v_i}t)$ $(i=1,3)$  and we neglect Bose enhancement and Pauli blocking in the final states.  The kinetic variables are assumed to obey a Lorentz-invariant regularization condition $S_2(s, t, u) = \theta(s\ge 2\mu_{D}^2)\theta(-s+\mu_{D}^2\le t\le -\mu_{D}^2)$ \cite{Jussi1} with the Debye screening mass squared $\mu_{D}^2 = (N_c + N_f/2 )g^2 T^2/3$. As we do not allow strong coupling constant to run in this study, we adopt a fixed strong coupling constant $\alpha_{\rm s}=g^{2}/4\pi$ which is set at 0.3 throughout this paper. The scattering rate for a hard parton of type $i$ with a thermal parton of type $j$ via a specific channel in the medium is,
\begin{equation}
\label{singlerate}
\Gamma_{ij\rightarrow kl} = \frac{1}{2E_1} \int \negthickspace \frac{d^3p_2}{(2\pi)^3 2E_2} \negthickspace \int \negthickspace \frac{d^3p_3}{(2\pi)^3 2E_3} \int \negthickspace \frac{d^3p_4}{(2\pi)^3 2E_4} \\
 f_j \, |M|_{ij\rightarrow kl}^2(s,t,u) \, S_2(s, t, u) \\
 (2\pi)^4 \delta^{(4)}(p_1+p_2-p_3-p_4).
\end{equation}
Summing over all types of incoming thermal parton $j(p_2)$ and all possible outgoing parton type pairs $k(p_3)$ and $l(p_4)$, the total scattering rate for an energetic parton $i(p_1)$) in a thermal medium is $\Gamma_i = \sum_{j(kl)} \, \Gamma_{ij\rightarrow kl}$.  The corresponding mean-free-path along a classical trajectory between adjacent scattering points is then $\lambda_i=1/\Gamma_i$. The probability of a parton-medium scattering in each time step $\Delta t$ is given by $P_{i}=1-e^{-\Gamma_i \Delta t}$. We use a time step $\Delta t$ that is much smaller than the mean-free-path to maximize the statistical accuracy of our simulations. Once we determine that there is a scattering in $\Delta t$, we sample every channel of the scattering processes using the partial rate of the single channel scattering as a weight. Both the jet shower partons $k(p_3)$ and the recoiled partons $l(p_4)$ after each scattering are tracked and will propagate further in the the medium. To take into account of the back reaction in the Boltzmann transport, the initial thermal parton $j(p_2)$ in each scattering process is denoted as ``negative'' a parton, which is also tracked and transported according to the Boltzmann equation. Contributions by these ``negative'' partons will be subtracted in the final parton spectra and jet energy, which essentially describes the depletion of the phase space of the initial thermal parton before each scattering. They are responsible for the wake excitation behind a propagating jet \cite{Li:2010ts}.  Notice that the incoming thermal partons and the outgoing recoiled partons appear in pairs. So both the recoiled partons and the ``negative'' partons should be taken into account when we study jet-induced medium excitations and compute the energy of reconstructed jets.

%\section{Results}
%\label{}

%%%%%%%%%%%%%%%%%%%%%%%%%%%%%%%%%%%%%%%%%%%%%%%%%%%%%%%%%%%%%%%%%%%%%

%%%%%%%%%%%%%%%%%%%%%%%%%%%%%%%%%%%%%%%%%%%%%%%%%%%%%%%%%%%%%%%%%%%%%
\begin{figure}[!ht]
\centerline{\includegraphics[width=7.5cm]{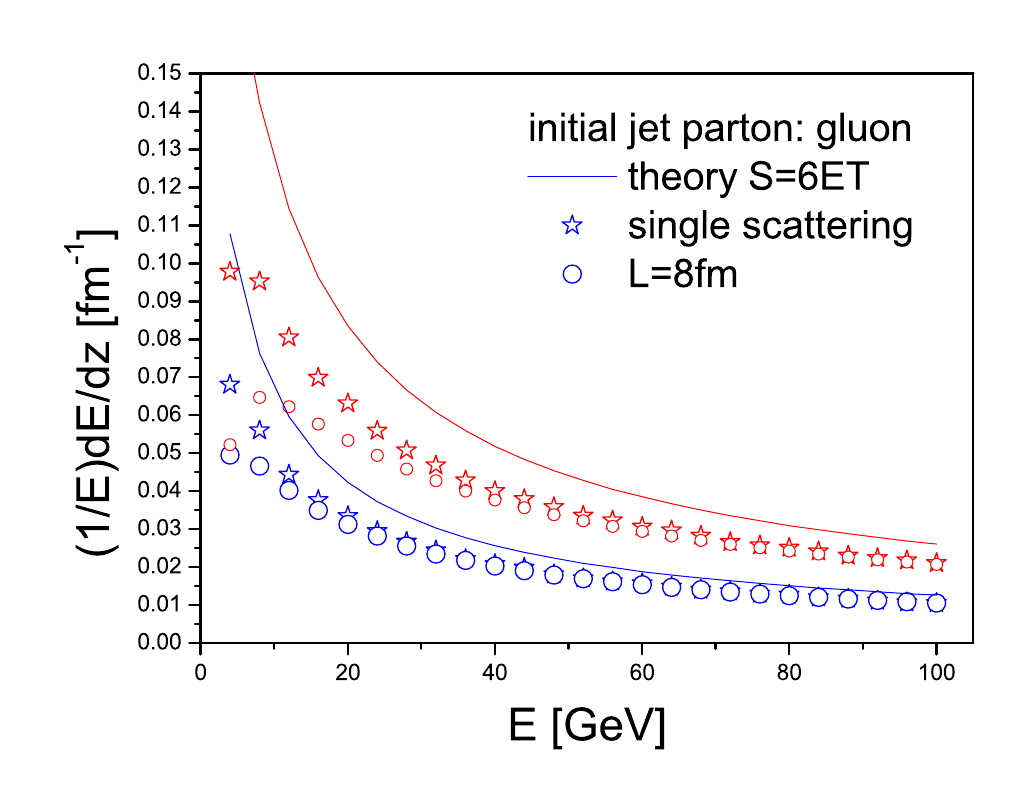}\includegraphics[width=7.5cm]{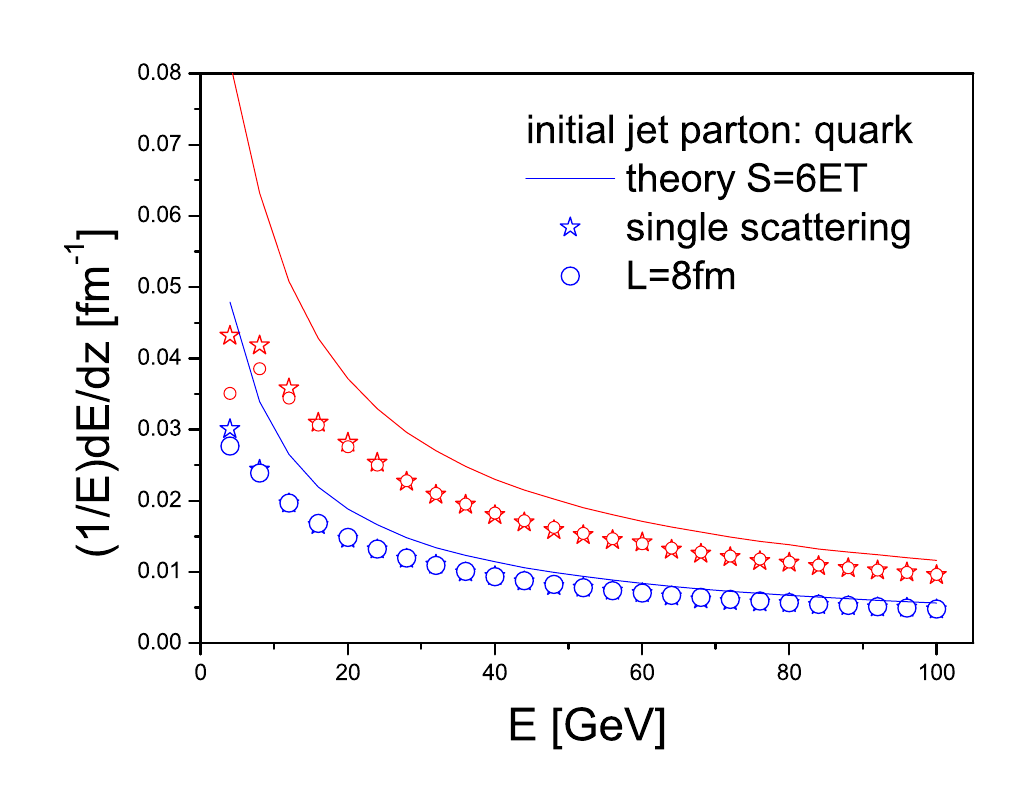}}
\caption{(Color online) Averaged fractional elastic energy loss per unit length $(dE/dz)/E$ of a gluon (left) and a quark (right) in a uniform and static QGP with temperature $T=200$ (bottom) and 300 MeV (top) as a function of the initial energy $E$ due to multiple scattering within a length $L=8$ fm (open circle) or a single scattering (open star) as compared to analytic results with a small angle approximation for the cross sections.}
\label{fig: elossE}
\end{figure}

%%%%%%%%%%%%%%%%%%%%%%%%%%%%%%%%%%%%%%%%%%%%%%%%%%%%%%%%%%%%%%%%%%%%%
\begin{figure}[!ht]
\centerline{\includegraphics[width=7.5cm]{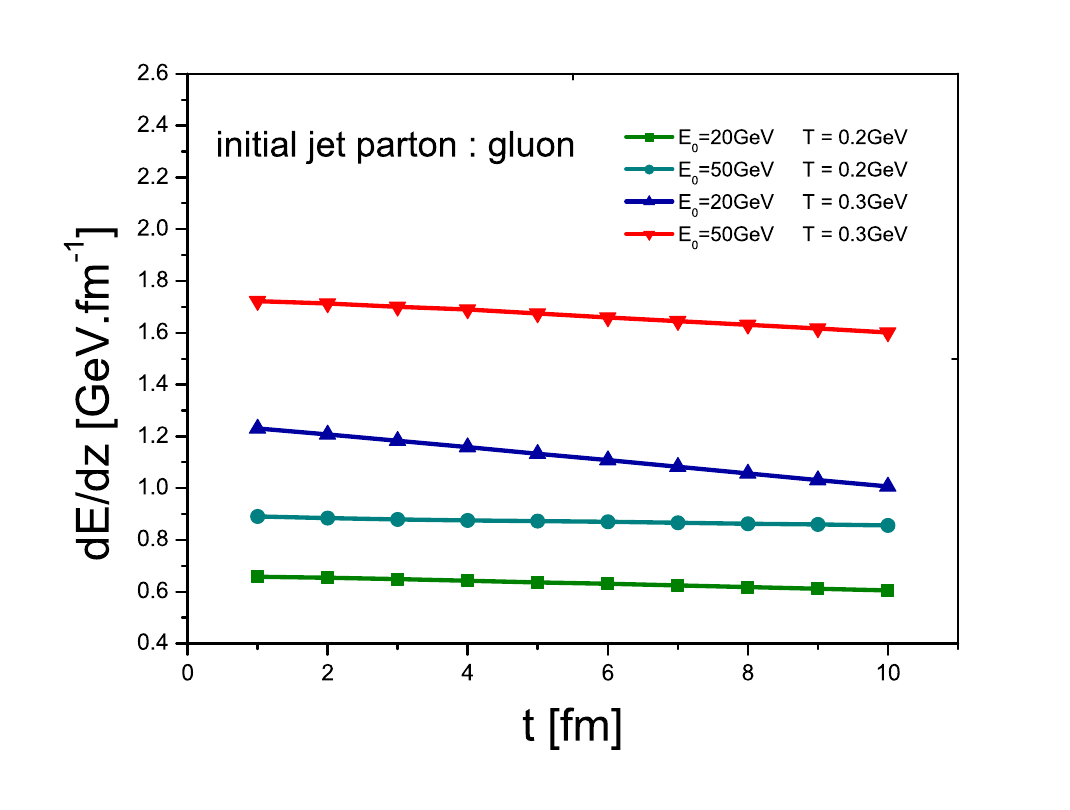}\includegraphics[width=7.5cm]{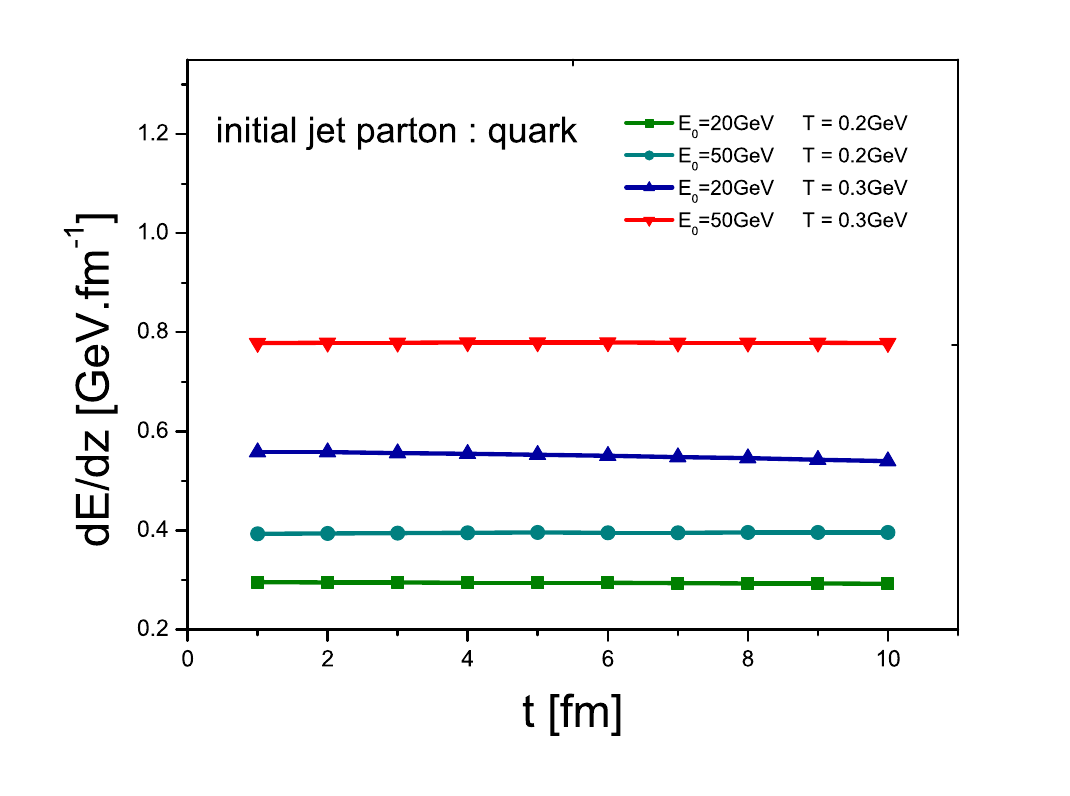}}
\caption{The elastic energy loss per unit path length $dE/dz$ of a gluon (left) and a quark (right) with different initial energy in a uniform and static QGP with temperature $T=200$ and 300 MeV as a function of the propagation time.}
\label{dedz}
\end{figure}

{\bf Results}. We first study the elastic energy loss within the LBT model with the complete set of $2\rightarrow 2$ processes.
Shown in Fig.~\ref{fig: elossE} is the fractional elastic energy loss per unit length of a gluon (left) and a quark (right) as a function of the initial parton energy in a static medium with a temperature $T=200$ MeV (lower)  and $T=300$ MeV (upper) and the total propagation length ($L=8$ fm) (open circle).
We also plot the average elastic energy loss from a single scattering (star). They are almost identical except for partons with small initial energy for which the energy-momentum conservation lowers the averaged energy loss per unit length of multiple scattering.  The difference is also bigger for gluons than quarks because gluon interaction cross section is a factor of 9/4 larger. Plotted as solid curves are the analytic estimates of the elastic energy loss assuming a small angle approximation for the cross sections \cite{xnwang},
\begin{equation}
\label{eloss}
\frac{dE_{\rm el}^{i}}{dz}=\frac{\Delta E}{\lambda }=C_{\rm el}^{i}2 \pi \alpha _{s}^{2} \, {{T}^{2}} \, \ln (\frac{s}{4\mu _{D}^{2}})
\end{equation}
where $C_{\rm el}^i$ is $9/4$ for a gluon and $1$ for a quark respectively and $s\approx 6ET$ is assumed.  As we can see, the analytic results from small angle approximation overestimate the elastic energy loss almost by a factor of 2 for partons with initial energy below 10 GeV. The agreement improves as the initial parton energy increases when the small angle scattering cross section becomes a better approximation.

In Fig.~\ref{dedz} we show the energy loss per unit path length $dE/dz$ of a gluon (left) and a quark (right) with different initial energy at different temperature as functions of propagation time. We can see that the averaged parton energy loss per unit path length of a parton going through multiple scattering decreases slightly with time (or length), especially for partons with smaller initial energy in a medium with a higher temperature. Such a decrease is also caused by energy-momentum conservation during multiple scattering. The time dependence for a quark is much weaker than a gluon because of the smaller interaction rate.

%%%%%%%%%%%%%%%%%%%%%%%%%%%%%%%%%%%%%%%%%%%%%%%%%%%%%%%%%%%%%%%%%%%%%

\begin{figure}[!ht]
\centerline{\includegraphics[width=7.5cm]{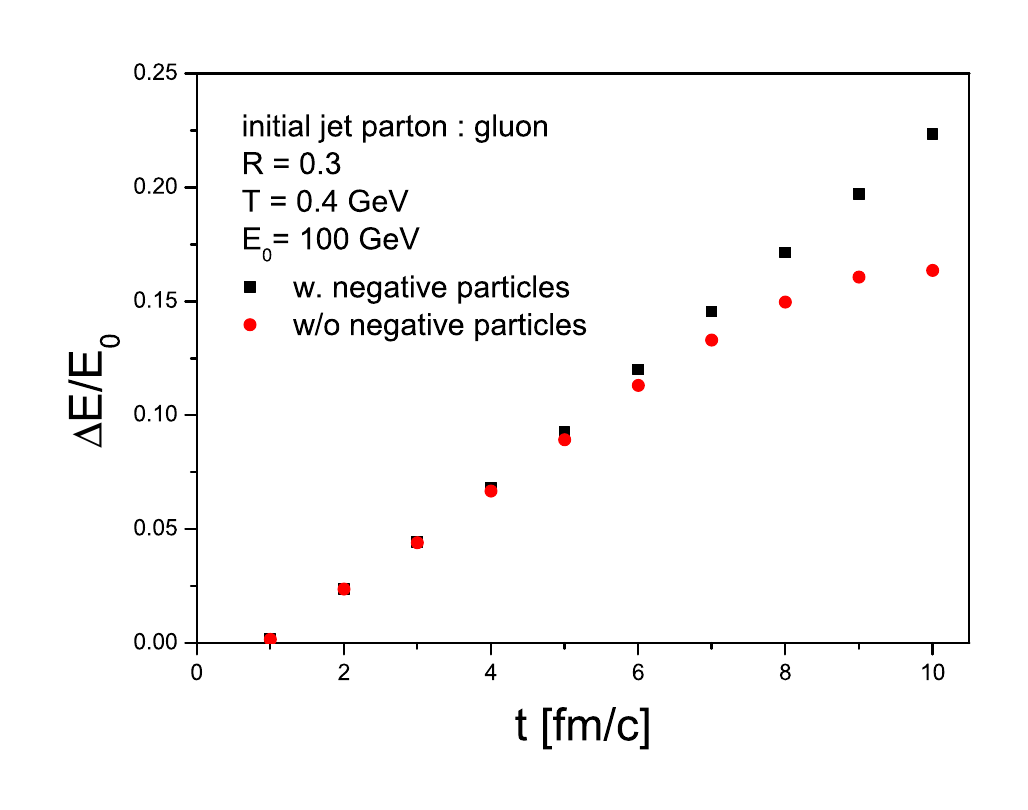}\includegraphics[width=7.5cm]{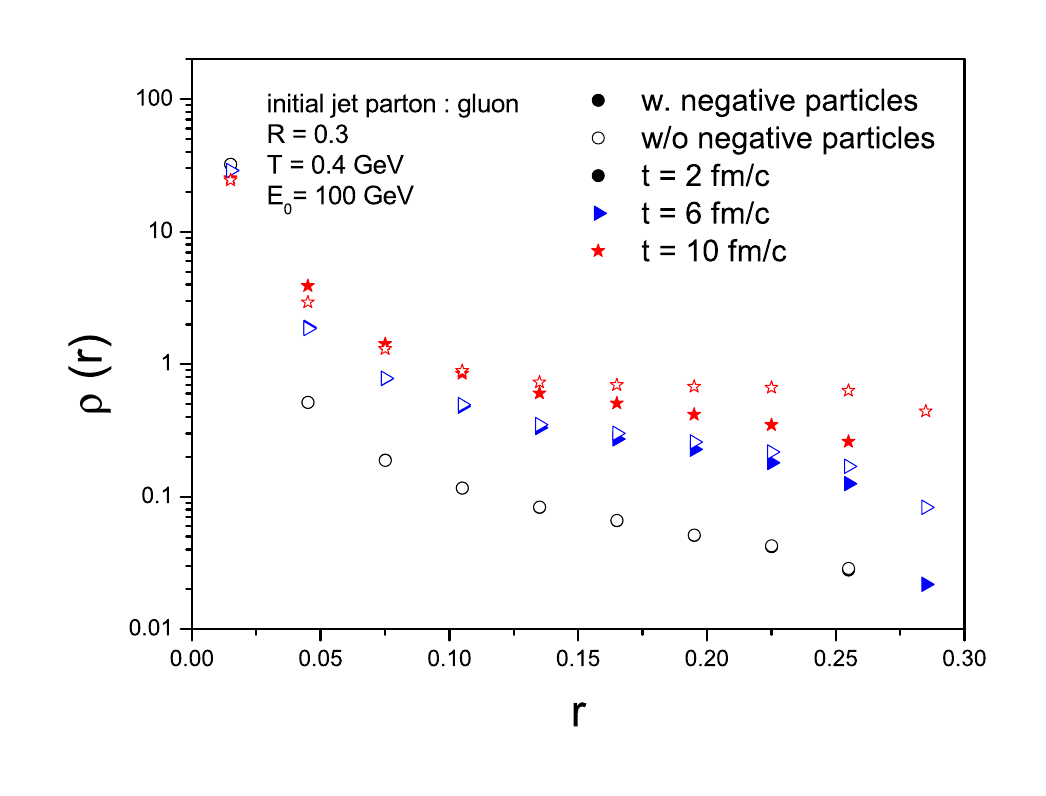}}
\caption{(Color online) (left) The fractional jet energy loss of the leading jet in a static and uniform QGP at a temperature $T=400$ MeV as a function of the propagation time with (solid square) and without (solid circle) contributions from ``negative" partons. (right) Jet transverse profiles of the leading jet at different times in a static medium with (solid symbols) and  without (open symbols) contributions from ``negative" partons.}
\label{fastjet}
\end{figure}

To study how the energy and transverse profile of a jet are modified by the interaction with the medium, we consider a gluon propagating in a uniform and static medium at temperature $T=400$ MeV. We set the initial energy of the gluon at $E=100$ GeV and use a jet-cone size $R=0.3$. Shown in Fig.~\ref{fastjet} (left) is the jet fractional energy loss of the leading jet as a function of propagation time. At the early stage of the propagation, the inclusion of  ``negative" particles or the jet-induced wake seems to make no difference. But at the later stage, the energy loss is much larger if we include these  ``negative" particles as the jet-induced wake becomes significant and depletes the energy inside the jet-cone at the late stage of the evolution. Most of these ``negative" partons are soft and in most of theoretical studies they are considered as part of the background. However, they are part of medium excitation induced by the propagating parton and should be considered as one calculate the energy and transverse profile of the reconstructed jets. The net jet energy loss has a linear dependence on the propagation time (or length).

The jet transverse profile is the average fraction of jet transverse momentum inside an annulus in the $\eta-\phi$ plane,
\begin{equation}
\rho(r)=\frac{1}{\Delta r}\frac{1}{N^\text{jet}}\sum_\text{jets}\frac{p_T (r-\Delta r/2,r+\Delta r/2)}{p_T (0,R)},
\end{equation}
where $p_T (r_1,r_2)$ is the summed $p_T$ of all partons in the annulus between radius $r_1$ and $r_2$ inside the jet-cone. It provides us with information on how the energy is distributed inside the jet cone and how it is modified by jet-medium interaction. Shown in Fig.~\ref{fastjet} (right) are the jet transverse profiles of the leading jet at different times.  We can see an obvious broadening of the leading jet transverse profile as it propagates through the medium because of the diffusion of the recoiled partons. However, this broadening is reduced at a later time when the energy of the soft ``negative" particles from the wake is subtracted since most of them are distributed toward the outer region of the jet cone.

This work is supported by the NSFC under Grant No. 11221504,  China MOST under Grant No. 2014DFG02050, the Major State Basic Research Development Program in China (No. 2014CB845404), U.S. DOE under Contract No. DE-AC02-05CH11231 and within the framework of the JET Collaboration.

%% The Appendices part is started with the command \appendix;
%% appendix sections are then done as normal sections
%% \appendix

%% \section{}
%% \label{}

%% References
%%
%% Following citation commands can be used in the body text:
%% Usage of \cite is as follows:
%%   \cite{key}         ==>>  [#]
%%   \cite[chap. 2]{key} ==>> [#, chap. 2]
%%

%% References with BibTeX database:

\bibliographystyle{elsarticle-num}
%\bibliography{<your-bib-database>}

%% Authors are advised to use a BibTeX database file for their reference list.
%% The provided style file elsarticle-num.bst formats references in the required Procedia style

%% For references without a BibTeX database:

\end{document}